\documentclass[aps,prl,twocolumn,superscriptaddress,nofootinbib]{revtex4-2}

\usepackage{graphicx}        
\usepackage{amsmath,amssymb} 
\usepackage{dcolumn}         
\usepackage{bm}              
\usepackage{hyperref}        
\usepackage[utf8]{inputenc}  
\usepackage{etoolbox}
\usepackage{orcidlink}

\begin{document}

\title{Energy gap of quantum spin glasses: a projection quantum Monte Carlo study}

\author{L. Brodoloni \orcidlink{0009-0002-0887-4020}}
\affiliation{CQM group, School of Science and Technology, Physics Division, Università di Camerino, 62032 Camerino, Italy}
\affiliation{INFN, Sezione di Perugia, I-06123 Perugia, Italy}

\author{G. E. Astrakharchik \orcidlink{0000-0003-0394-8094}}
\affiliation{Departament de Física, Universitat Politècnica de Catalunya, Campus Nord B4-B5, E-08034 Barcelona, Spain}

\author{S. Giorgini \orcidlink{0000-0001-9146-7025}}
\affiliation{Pitaevskii BEC Center, CNR-INO and Dipartimento di Fisica, Universit\`a di Trento, I-38123 Trento, Italy}

\author{S. Pilati \orcidlink{0000-0002-4845-6299}}
\affiliation{CQM group, School of Science and Technology, Physics Division, Università di Camerino, 62032 Camerino, Italy}
\affiliation{INFN, Sezione di Perugia, I-06123 Perugia, Italy}
    

\begin{abstract}
The performance of quantum annealing for combinatorial optimization is fundamentally limited by the minimum energy gap $\Delta$ encountered at quantum phase transitions. We investigate the scaling of $\Delta$ with system size $N$ for two paradigmatic quantum spin-glass models: the two-dimensional Edwards-Anderson (2D-EA) and the all-to-all Sherrington-Kirkpatrick (SK) models. Utilizing a newly proposed unbiased energy-gap estimator for continuous-time projection quantum Monte Carlo simulations, complemented by high-performance sparse eigenvalue solvers, we characterize the gap distributions across disorder realizations. It is found that, in the 2D-EA case, the inverse-gap distribution develops a fat tail with infinite variance as $N$ increases. This indicates that the unfavorable super-algebraic scaling of $\Delta$, recently reported for binary couplings [Nature {\bf 631}, 749 (2024)], persists for the Gaussian disorder considered here, pointing to a universal feature of 2D spin glasses. Conversely, the SK model retains a finite-variance distribution, with the disorder-averaged gap following a rather slow power law, close to $\Delta \propto N^{-1/3}$. 
This finding provides a promising outlook for the potential efficiency of quantum annealers for optimization problems with dense connectivity.
\end{abstract}

\maketitle

\emph{Introduction.}
The spectral gap $\Delta$ -- the energy difference between the ground state and the first excited state -- is a fundamental quantity across condensed-matter physics, quantum chemistry, and quantum information science. In the context of adiabatic quantum computing, the smallest value $\Delta$ attains on the annealing path governs the computational complexity of solving combinatorial optimization problems. In fact, the time required to avoid non-adiabatic transitions and ensure an optimal solution scales as $1/\Delta^2$ ~\cite{RevModPhys.90.015002}. 
As many quadratic optimization problems can be mapped onto Ising spin glasses, significant effort has focused on these models and, in particular, on the spin-glass quantum phase transition, where $\Delta$ is expected to vanish rapidly with the number of spins $N$~\cite{Kazutaka2010,BAPST2013127,PRXQuantum.5.020313,PhysRevLett.101.170503,tang2025stretchedexponentialscalingparityrestricted}.
Despite its importance, estimation of the energy gaps remains a formidable computational problem~\cite{6875473,10.1561/0400000066,Gharibian2019complexityof}, leading to conflicting reports on the scaling of $\Delta$ in the literature~\cite{PhysRevLett.72.4141,PhysRevLett.72.4137,PhysRevE.96.022139,PhysRevE.87.032154,PhysRevB.94.024201}. 
This uncertainty has spurred the search for more efficient numerical algorithms capable of probing the large $N$ regime~\cite{PhysRevE.85.036705,PhysRevLett.121.167204,PhysRevResearch.6.033322,doi:10.1126/sciadv.abl6850}.

Recent large-scale path-integral Monte Carlo (PIMC) simulations of the two-dimensional quantum Edwards-Anderson (2D-EA) model~\cite{Bernaschi2024} have clarified the role of the parity symmetry. 
It was found that the smallest gap -- separating the even ground state from the first odd excited state -- exhibits a super-algebraic scaling with system size. 
Instead, for the symmetry restricted gap, a power-law scaling was found, indicating a polynomial computational complexity. 
However, these findings were restricted to square lattices with binary couplings, whereas many realistic optimization problems -- e.g., mean-variance portfolio optimization or number partitioning problems -- involve denser connectivity and continuous coupling distributions~\cite{10.3389fphy.2014.00005,Venturelli2019}.

In this Letter, we investigate the scaling of $\Delta$ for both the 2D-EA and the all-to-all Sherrington-Kirkpatrick (SK) models featuring Gaussian couplings and uniform transverse field. Our analysis is powered by unbiased continuous-time projection quantum Monte Carlo (PQMC) simulations~\cite{becca2017quantum}. We show that the energy-gap estimator is unbiased  -- meaning it is independent of the choice of the guiding wave function -- ensuring high-fidelity results for systems up to $N \gtrsim 100$. These simulations are complemented by high-performance sparse eigenvalue solvers for smaller sizes ($N \lesssim 30$).

Our results demonstrate that, for the 2D-EA model, the distribution of the inverse gap $\eta = 1/\Delta$ acquires a fat tail with infinite variance for a finite $N$. This confirms that the super-algebraic scaling observed in Ref.~\cite{Bernaschi2024} for the symmetry-unrestricted gap is a universal feature that persists for continuous couplings. Conversely, for the SK model, the variance of $\eta$ remains finite, and the disorder-averaged gap follows a favorable power-law scaling with a rather small exponent, close to $\Delta \propto N^{-1/3}$. This suggests that the all-to-all connectivity of the SK model provides a significant advantage for quantum annealing over optimization problems that can be mapped to 2D sparse graphs.

\emph{Models and methods.}
The quantum Ising Hamiltonians we address are defined using conventional Pauli matrices $\sigma^x_{i}$ and $\sigma^z_{i}$, as
\begin{equation}
\hat{H}=-\Gamma \sum_{i=1}^{N} {\sigma}^{x}_{i} -\sum_{\left<i,j\right>} J_{ij}{\sigma}^{z}_{i} {\sigma}^{z}_{j}.
\label{H}
\end{equation}
The first term describes the effect of a uniform transverse field $\Gamma$. The second term accounts for interactions which, in the 2D-EA model, act only on the four nearest neighbors of a square lattice of side length $L=\sqrt{N}$, obeying periodic boundary conditions. The random couplings are sampled from a normal distribution with zero mean and variance $J^2$, namely $J_{ij} \sim \mathcal{N}(0, J^2)$. In the SK model, the interactions act on every spin pair with couplings $J_{ij} \sim \mathcal{N}(0, J^2/N)$. 
In the following, the scale $J$ is used as the energy unit.

We implement the PQMC algorithm in the standard computational basis of eigenstates $|{\mathbf x} \rangle$ of $\sigma^z_i$, such that $\sigma^z_i| {\mathbf x} \rangle = x_i | {\mathbf x} \rangle$, and the $2^N$ spin configurations are denoted as ${\mathbf x} = (x_1,\dots,x_N)$, with $x_i=\pm1$ for $i=1,\dots,N$.
Instead of directly solving the imaginary-time Schrödinger equation for the many-body wave function $\psi({\mathbf x},\tau) = \langle x | \psi(\tau)\rangle$, we improve the convergence by performing the importance sampling with a guiding wave function $\psi_g({\mathbf x})=\langle x | \psi_g\rangle$. 
The population of walkers then samples configurations proportionally to the 
distribution $f({\mathbf x},\tau)=\psi_g({\mathbf x})\psi({\mathbf x},\tau)$, which evolves according to the imaginary-time ($\tau=it/\hbar$) Schrödinger equation, modified by the guiding function and a reference energy $E_{\mathrm{ref}}$. 
The corresponding Green's function is
\begin{equation}
G({\mathbf x},{\mathbf x}_0,\tau)=\psi_g({\bf x})\langle{\mathbf x}|e^{-\tau \hat{H}_r}|{\mathbf x}_0\rangle \psi_g^{-1}({\bf x}_0),
\label{eqG}
\end{equation}
where $\hat{H}_r =\hat{H} - E_{\mathrm{ref}}$ is the energy-shifted Hamiltonian.
The Hamiltonian~(\ref{H}) commutes with the parity operator $\hat{P}=\prod_{i=1}^N\sigma^x_i$, whose eigenvalues are either $-1$ or $1$. Thus, the eigenstates separate into the odd and even sectors, with corresponding energy levels $E_{o,0}$, $E_{o,1}$, $\dots$, and  $E_{e,0}$, $E_{e,1}$, $\dots$, respectively.
The states $|\psi_0\rangle$ and $|\psi_g\rangle$ are even, and the ground-state energy is $E_0 = E_{e,0}$.

To derive the gap estimator, we evaluate the matrix element $\langle\psi_g|\hat{O}(\tau)\hat{O}(0)|\psi_0\rangle=\langle\psi_g|\hat{O}(0)\hat{O}(-\tau)|\psi_0\rangle$, where the equality follows from the time invariance. 
Here, $\hat{O}(\tau)$ denotes the Heisenberg representation of the observable $\hat O$ at imaginary time $\tau$, defined as $\hat{O}(\tau)=e^{\tau\hat{H}_r}\hat{O}e^{-\tau\hat{H}_r}$.
After equilibration from the initial state, $E_{\mathrm{ref}}$ provides an unbiased estimate of the ground-state energy $E_0$. 
We can hence write $e^{\tau\hat{H}_r}|\psi_0\rangle = |\psi_0\rangle$ and, using two completeness relations, the matrix element can be written as
\begin{equation}
\langle\psi_g|\hat{O}\hat{O}(-\tau)|\psi_0\rangle
=
\sum_{{\mathbf x},{\mathbf x}_0}  \psi_g({\bf x})\langle{\mathbf x}|\hat{O}e^{-\tau H_r}\hat{O}|{\mathbf x}_0\rangle \psi_0({\mathbf x}_0) \,.
\nonumber
\end{equation}
For diagonal operators, \emph{i.e.} such that $\hat{O}|{\mathbf x}\rangle=O(\mathbf x)|{\mathbf x}\rangle$, the above equation can be recast as
\begin{equation}
\langle\psi_g|\hat{O}\hat{O}(\!-\tau\!)|\psi_0\rangle\!\!
=\!\!\!
\sum_{{\mathbf x},{\mathbf x}_0}\!\!O(\mathbf x)O({\mathbf x}_0) \psi_g({\bf x})
\langle{\mathbf x}|e^{-\tau \hat{H}_r}\!|{\mathbf x}_0\rangle \psi_0({\mathbf x}_0) \,.
\nonumber
\end{equation}
Recalling that after equilibration for any $\tau$ one has $f({\mathbf x},\tau)=\psi_g(\mathbf{x})\psi_0(\mathbf{x})$ and using the definition of
$G({\mathbf x},{\mathbf x}_0,\tau)$ in Eq.~\eqref{eqG},
one gets
\begin{align}
\langle\psi_g|\hat{O}\hat{O}(-\tau)|\psi_0\rangle
& =
\sum_{{\mathbf x},{\mathbf x}_0} O({\mathbf x})O({\mathbf x}_0) 
G({\mathbf x},{\mathbf x}_0,\tau)f({\mathbf x_0},0)\nonumber\\
& = \sum_{{\mathbf x}} O({\mathbf x})O({\mathbf x}_0) 
f({\mathbf x},\tau)
\;.
\label{MCcorrelator}
\end{align}
The right-hand side in the last line is easily stochastically estimated in the PQMC simulation by allowing walkers to inherit the observable values at previous times and  averaging over all configurations ${\mathbf x}$ present at imaginary time $\tau$ that were in ${\mathbf x}_0$ at the earlier time $\tau=0$.

Now, for an odd operator, such that $\hat{P}\hat{O}\hat{P}^{\dagger}=-\hat{O}$, one can write down the expansions 
$\hat{O}|\psi_g\rangle= \sum_{n_o} d_{n_o} |\psi_{n_o}\rangle$ and $\hat{O}|\psi_0\rangle= \sum_{n_o} c_{n_o} |\psi_{n_o}\rangle$, where the index $n_o=0,1,\dots$ labels the odd eigenstates. 
Using these expansions, the matrix element can be rewritten as
\begin{align*}
\langle\psi_g|\hat{O}\hat{O}(-\tau)|\psi_0\rangle 
&=\sum_{n_o}c_{n_o}d^*_{n_o}e^{-(E_{n_o}-E_0)\tau}\\
& \simeq c_{n_o}d^*_{n_o}e^{-(E_{o,0}-E_0)\tau} \;,
\label{eq6}
\end{align*}
where the last result holds for long enough $\tau$ so that higher (odd) excited states do not contribute.
Hence, the long-time exponential decay of the matrix element allows one to determine the lowest odd energy gap $\Delta_{o,1}=E_{o,0}-E_{0}$. 
Importantly, the choice of $\psi_g(\mathbf{x})$ does not affect the gap estimate, meaning that we have derived a pure estimator. Yet, a good approximation $\psi_g(\mathbf{x})\simeq\psi_0(\mathbf{x})$ improves the statistics and eliminates the walker population control bias~\cite{PhysRevE.101.063308,PhysRevE.100.043301}. 
Here, we consider neural quantum states in the form of restricted Boltzmann machines~\cite{doi:10.1126/science.aag2302} featuring (unless otherwise specified) $N_h = N$ hidden neurons. These states are optimized via variational energy minimization using the NetKet library~\cite{CARLEO2019100311,netket3:2022}.

An even operator, such that $\hat{P}\hat{O}\hat{P}^{\dagger}=\hat{O}$, allows extracting the even gap $\Delta_{e,0}=E_{e,1}-E_{0}$ from the exponential decay to a finite asymptotic value.
Demonstrative benchmarks for both odd and even operators are provided in the Supplemental Material~\cite{SM}. Details on the choice of  the operators and on the fitting protocols are reported in the End Matter.
Estimating $\Delta_{e,0}$ via PQMC simulations turns out to be  more computationally demanding than $\Delta_{o,0}$, since the successive gap $\Delta_{e,1}$ is often of similar (large) magnitude. This implies that higher-level contributions to the imaginary-time correlation vanish in regimes where the signal-to-noise ratio is small.
Hence, in this case we resort to exact eigenvalue calculations based on the Lanczos algorithm. Specifically, we adopt the implementation provided by the sparse linear algebra library Cupy~\cite{nishino2017cupy} for graphic processing units, exploiting the QuSpin library for efficient handling of large sparse Hamiltonian matrices~\cite{10.21468/SciPostPhys.7.2.020}.
Due to parity symmetry, block diagonalization of the odd and even sectors is possible.
%
Conversely, for most disorder realizations one has $\Delta_{o,0}\ll \Delta_{o,1}$, allowing a computationally efficient estimation of the odd gap via PQMC simulations for sufficiently many disorder realizations and sizes $N\lesssim 125$.

\begin{figure}[h]
\centering
\includegraphics[width=1.\columnwidth]{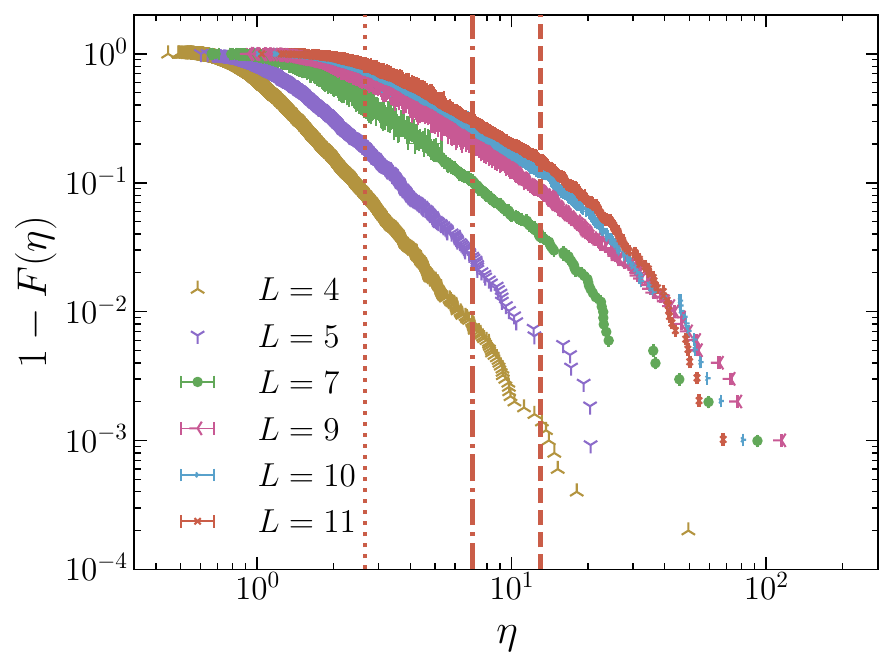}
\caption{
Complementary empirical distribution function $1-F(\eta)$ of the inverse odd gap $\eta=1/\Delta$ of the Gaussian 2D-EA model at the spin-glass quantum phase transition.
The various symbols correspond to the lattice sizes $L=\sqrt{N}$ reported in the legend.
The three vertical lines indicate the smallest $\eta$ accounted for in the Hill estimator for $L=11$, for three sample fractions: $\kappa=0.8$, $\kappa=0.3$, and $\kappa=0.15$ (from left to right).
}
\label{fig1}
\end{figure}

\begin{figure}[h]
\centering
\includegraphics[width=1.\columnwidth]{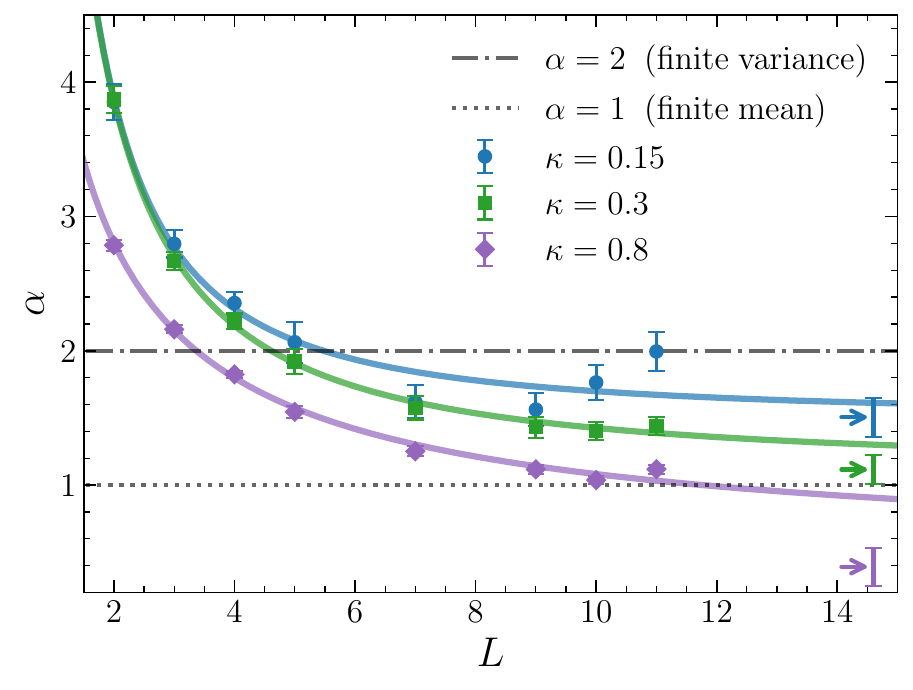}
\caption{
Hill estimator for the tail index $\alpha$ (see Eq.~\eqref{hill}) for the Gaussian 2D-EA Hamiltonian as a function of the lattice size $L$.
The three symbols correspond to the sampled fractions $\kappa$ reported in the legend. 
The continuous curves represent power-law fitting functions. The error bars are determined via a bootstrap method with $N_{\mathrm{boot}}=2000$ resamples.
The arrows on the right point to the extrapolation $L\rightarrow \infty$.
The two horizontal straight lines indicate the thresholds for finite mean and variance.
}
\label{fig2}
\end{figure}

\emph{Results.}
A preliminary analysis based on the exact eigenvalue solver reveals that, for the  quantum spin models addressed in this article, in the critical regime the lowest odd energy gap $\Delta_{o} \equiv\Delta_{o,0}$ is smaller than the lowest even energy gap $\Delta_{e} \equiv\Delta_{e,0}$, for all considered disorder realizations. Hence, hereafter, the smallest energy gap $\Delta$, which is the main focus of this article, is identified with the former, $\Delta \equiv \Delta_{o}$. Hence, an odd operator $\hat{O}$ is adopted in the PQMC gap-estimator described above.

We first focus on the Gaussian 2D-EA model, tuning the transverse field at $\Gamma=1.98$, corresponding to the estimated critical point of the spin-glass quantum phase transition $\Gamma = 1.98(7)$~\cite{PhysRevE.110.065305}. Since the annealing times scale with the square of the inverse gap $\eta$, it is interesting to analyze the distribution of this latter quantity across disorder realizations.
The sample size ranges from $N_r \approx 5000$ disorder realizations in the case of eigenvalue-solver calculations for $N\approx 20$, to $N_r = 1000$ in PQMC simulations for $N \simeq 120$.
Fig.~\ref{fig1} shows the complementary empirical distribution function $1-F(\eta)$, namely, the estimate of the probability of finding inverse gaps larger than  a given threshold $\eta$. As expected, this probability decreases with $\eta$, but the slope rapidly diminishes with the lattice length $L$.
To quantify this trend, we consider the Hill estimator for the tail index $\alpha$, defined as
\begin{equation}
    \frac{1}{\alpha} = \frac{1}{k} \sum_{i=1}^{k} \ln \left( \frac{\eta_{(N_r-i+1)}}{\eta_{(N_r-k)}} \right)\, ,
    \label{hill}
\end{equation}
where the $N_r$ realizations $\eta_1$, $\eta_2$, $\dots$, $\eta_{N_r}$ are arranged in ascending order,  {\it i.e.} $\eta_{N_r}>\eta_{N_r-1}>\dots$, and the $k<N_r$ largest values are considered.
Note that this estimator assumes a power-law tail of the probability density function of the form $f(\eta) = F^{\prime}(\eta) \propto \eta^{-(\alpha+1)}$ for large $\eta$.
We first consider a rather large sample fraction $\kappa = 0.80 \simeq k/N_r $, such that the whole bulk regime of the function $1-F(\eta)$ -- beyond the shoulder at small $\eta$ -- is accounted for.  As shown in Fig.~\ref{fig2}, the resulting tail index is relatively small, namely $\alpha \approx 2$, already for small sizes $L\simeq  4$, denoting a distribution with a rather fat tail. Notably, $\alpha$ further decreases with the system size $L$. In fact, if an empirical power-law fitting function is assumed, the threshold $\alpha=1$, below which the disorder ensemble average $[\eta]$ is diverging, is crossed for finite sizes $L\simeq 12$. The extrapolation to $L\rightarrow \infty$ is sizably below the $\alpha=1$ threshold.
The same conclusion is reached also by directly performing power-law fits in the bulk region of $1-F(\eta)$ (see End Matter). This finding is consistent with the finite-temperature PIMC simulations of the binary 2D-EA model of Ref.~\cite{Bernaschi2024}.
For the feasible sizes, the PQMC simulations also reach the smallest gaps, allowing us to focus in more detail on the large $\eta$ regime of the empirical distribution function.
Considering smaller sample fractions $\kappa \lesssim 0.3$, such that only the large $\eta$ regime is accounted for, the extrapolated tail index appears to saturate to somewhat larger values, in the range $1\leq \alpha \leq 2$. This range corresponds to a distribution with infinite variance but finite average. Notice that an infinite variance precludes estimating the distribution average from a finite sample.
Concerning
the potential efficiency of quantum annealers, tail indices $\alpha \lesssim 2$ still signal a large probability that a random instance requires problematically long annealing times.

\begin{figure}[h]
\centering
\includegraphics[width=1.\columnwidth]{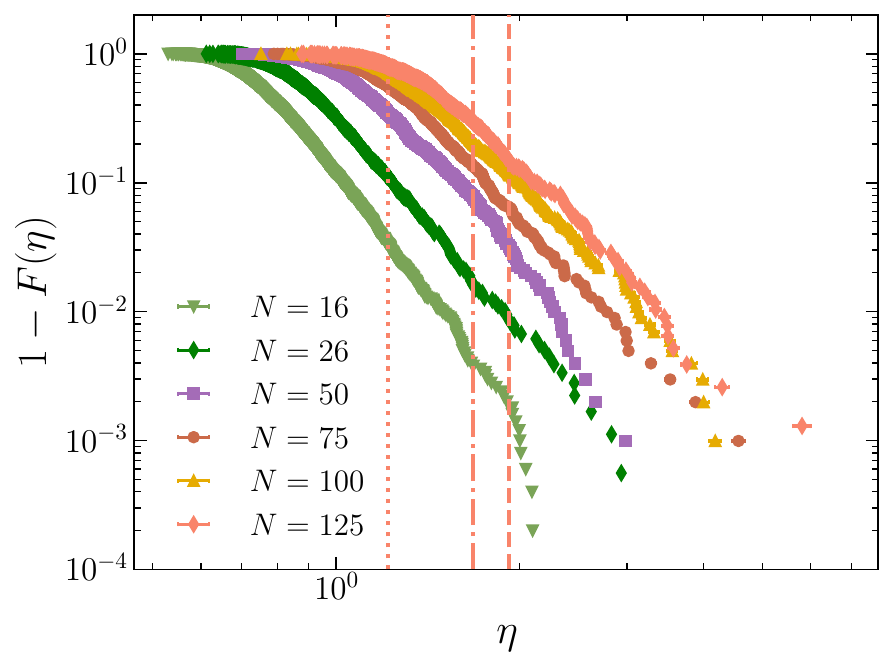}
\caption{
Complementary empirical distribution function $1-F(\eta)$ for the SK model. Symbols and lines are defined as in Fig.~\ref{fig1}.
}
\label{fig3}
\end{figure}

\begin{figure}[h]
\centering
\includegraphics[width=1.\columnwidth]{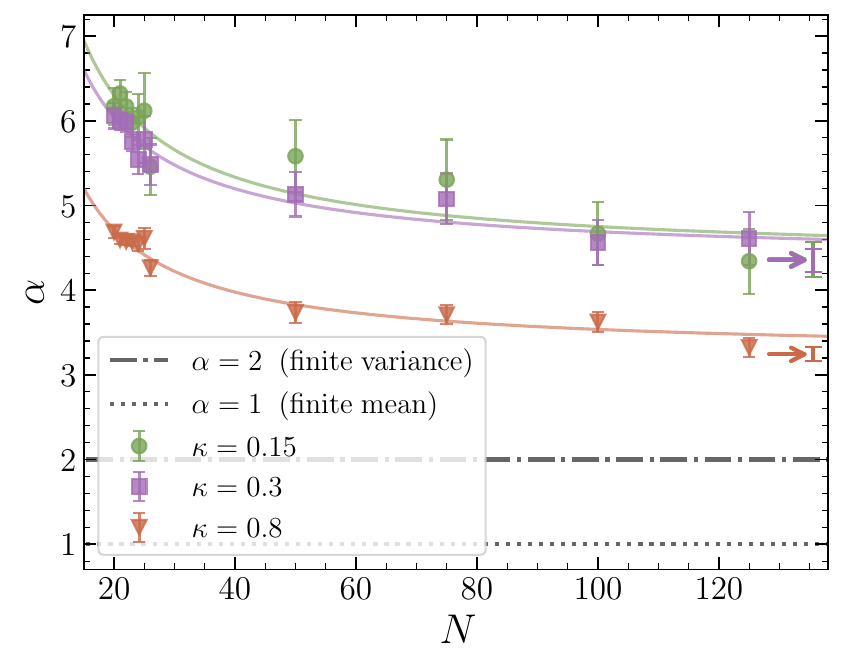}
\caption{
Hill estimator $\alpha$ as a function of the number of spins $N$ for the SK Hamiltonian. Symbols and lines are defined as in Fig.~\ref{fig3}.
}
\label{fig4}
\end{figure}

\begin{figure}[h]
\centering
\includegraphics[width=1.\columnwidth]{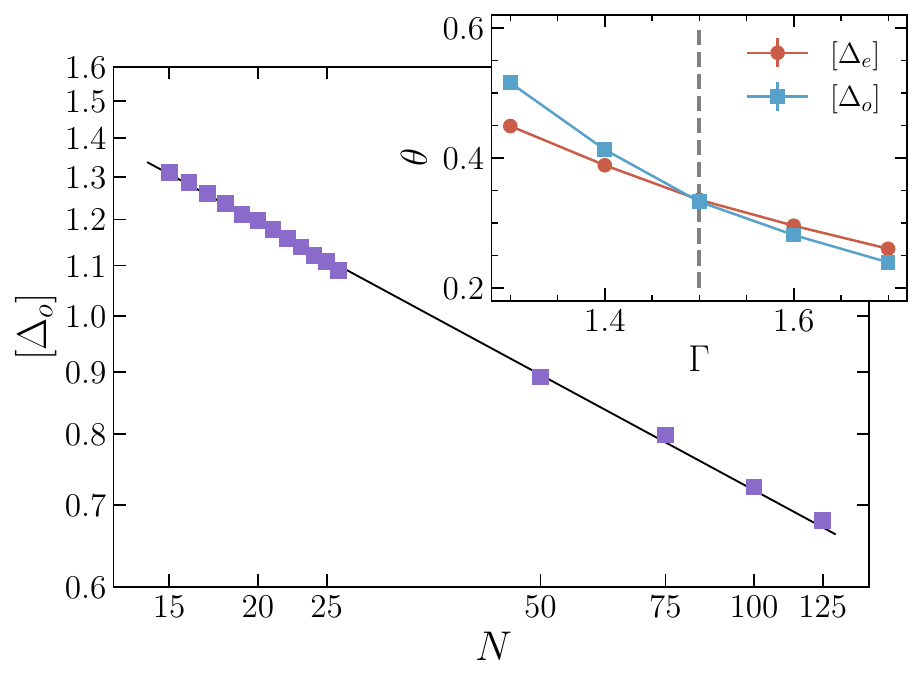}
\caption{
Main panel: Odd energy gap $\Delta_o$, averaged over disorder realizations, as a function of the number of spins $N$, for the SK model.
The black line represents a power-law  fit $\Delta = C / N^\theta$, providing $C=3.06(3)$ and $\theta=0.32(1)$.
Inset: 
Scaling exponent $\theta$ as a function of the transverse field $\Gamma$ for the odd gap $\Delta_o$ and the even gap $\Delta_e$.
The dashed vertical line shows the position of the critical point.
}
\label{fig5}
\end{figure}

Since typical combinatorial optimization problems often lead to a much denser connectivity than the four nearest neighbors of the square lattice, it makes sense to address the opposite regime, considering the all-to-all connectivity of the quantum SK Hamiltonian.
Its inverse-gap distribution at the spin-glass quantum critical point $\Gamma \simeq 1.5$~\cite{PhysRevLett.70.3147,DavidLancaster1997,PhysRevE.78.061121,PhysRevE.96.032112,kiss2023exact,PhysRevLett.129.220401} is analyzed in Fig.~\ref{fig3}. 
In this case, the decay rate of $1-F(\eta)$ at large $\eta$ is less sensitive to the system size $N$. Indeed, for sizes $N\rightarrow \infty$, the estimated tail index saturates well above the threshold $\alpha=2$ (see Fig.~\ref{fig4}). Hence, the disorder averaged inverse gap $\left[\eta\right]$ is well defined for all finite sizes, and a stochastic estimate must converge for large sample size. 
The scaling of $\left[\eta\right]$ with $N$ is analyzed in the Supplemental Material~\cite{SM}.
In Fig.~\ref{fig5}, the more common scaling of the disorder averaged  gap $\left[\Delta\right]$ versus $N$ is shown. Importantly, it is found that at the critical point a power-law fit $\left[\Delta\right]\propto N^{-\theta}$ with $\theta = 0.32(1)$ accurately describes both the eigenvalue-solver and the PQMC data. 
It is worth mentioning that a $1/3$ exponent was discussed for the two-pattern Gaussian Hopfield model~\cite{knysh2015computational}, and argued to apply also to other mean-field models.
On the other hand, a Hartree-Fock theory for the quantum SK model predicted $\theta \simeq 0.66$~\cite{PhysRevB.93.134202}.
Other studies addressed the classical SK model and the scaling of the free-energy barrier height~\cite{Bittner2006,PhysRevE.105.034138,doi:10.1073/pnas.1910936117}, finding a power-law growth with exponent $1/3$.

The inset of Fig.~\ref{fig5} compares the scaling exponents for odd and even gaps. For the latter, and for the odd gap at transverse fields other than $\Gamma=1.5$, eigenvalue-solver data up to sizes $N=26$ are considered. Assuming a power-law scaling even slightly away from the critical point, we find that the odd gap exponent is larger than the even gap exponent (slightly) within the spin-glass phase $\Gamma \lesssim 1.5$, and vice versa in the paramagnetic phase. Interestingly, the two exponents agree within statistical uncertainties at the critical point. This indicates that both symmetry-changing and symmetry-restricted excitations are suppressed by a power-law scaling gap.

\emph{Discussion.}
In summary, our investigation of the quantum 2D-EA model demonstrates that the super-algebraic scaling of the symmetry unrestricted (i.e., odd) energy gap -- previously observed for binary disorder via finite-temperature PIMC simulations -- is a universal feature that persists for Gaussian couplings and continuous imaginary time. In contrast, the quantum SK model exhibits a more benign power-law scaling with a favorable exponent $\theta \simeq 1/3$, which remains stable at least up to $N \simeq 125$.
This finding offers a promising outlook for quantum annealing of combinatorial optimization problems featuring a dense connectivity.
While our focus has been on the quantum critical point, we acknowledge that the critical region is not the sole bottleneck in quantum optimization. Exponentially small gaps can emerge deep within the spin-glass phase~\cite{knysh2015computational}. Future work should bridge the gap between these paradigms by examining systems with intermediate connectivity, such as power-law random interactions relevant to Rydberg-atom arrays~\cite{r6dq-2ztr} or trapped ions~\cite{Smith2016}.

Finally, this study relied on the implementation of a continuous-time PQMC algorithm with a pure gap estimator. As we have shown, this estimator is independent of the guiding wave function  and it provides  a robust and versatile tool  to be used in many-body problems. We expect this methodology to be instrumental in resolving spectral properties across diverse fields, from frustrated magnetism to quantum chemistry.
The data presented in this article are available on-line~\cite{brodoloni_2026_18705683}.

\begin{acknowledgments}
We thank M. Murdaca and E. Vitali for assistance on the LUMI supercomputer.
Support from the following sources is acknowledged: 
PNRR MUR project PE0000023-NQSTI; 
PRIN 2022 MUR project ``Hybrid algorithms for quantum simulators'' -- 2022H77XB7; 
PRIN-PNRR 2022 MUR project ``UEFA'' -- P2022NMBAJ; 
National Centre for HPC, Big Data and Quantum Computing (ICSC), CN00000013 Spoke 7 -- Materials \& Molecular Sciences; 
CINECA award INF25\_lincoln; 
EuroHPC Joint Undertaking for awarding access to the EuroHPC supercomputer LUMI, hosted by CSC (Finland), through EuroHPC Development and Regular Access calls.
\end{acknowledgments}

\bibliography{references}

@book{suzuki2012quantum,
  title={Quantum {Ising} phases and transitions in transverse {Ising} models},
  author={Suzuki, Sei and Inoue, Jun-ichi and Chakrabarti, Bikas K},
  volume={862},
  year={2012},
  publisher={Springer Berlin, Heidelberg}
}

@article{PhysRevB.53.8486,
  title = {Numerical study of the random transverse-field {Ising} spin chain},
  author = {Young, A. P. and Rieger, H.},
  journal = {Phys. Rev. B},
  volume = {53},
  issue = {13},
  pages = {8486--8498},
  numpages = {0},
  year = {1996},
  month = {Apr},
  publisher = {American Physical Society},
  doi = {10.1103/PhysRevB.53.8486},
  url = {https://link.aps.org/doi/10.1103/PhysRevB.53.8486}
}

@Article{10.21468/SciPostPhysLectNotes.82,
	title={{The quantum Ising chain for beginners}},
	author={Glen Bigan Mbeng and Angelo Russomanno and Giuseppe E. Santoro},
	journal={SciPost Phys. Lect. Notes},
	pages={82},
	year={2024},
	publisher={SciPost},
	doi={10.21468/SciPostPhysLectNotes.82},
	url={https://scipost.org/10.21468/SciPostPhysLectNotes.82},
}

@article{tang2025stretchedexponentialscalingparityrestricted,
      title={Stretched Exponential Scaling of Parity-Restricted Energy Gaps in a Random Transverse-Field {Ising} Model}, 
      author={G. -X. Tang and J. -Z. Zhuang and L. -M. Duan and Y. -K. Wu},
      year={2025},
      journal={arXiv:2512.03526},
      doi={10.48550/arXiv.2512.03526},
      url={https://arxiv.org/abs/2512.03526}, 
}

@article{PFEUTY197079,
title = {The one-dimensional {Ising} model with a transverse field},
journal = {Ann. Phys. (N. Y.)},
volume = {57},
number = {1},
pages = {79-90},
year = {1970},
issn = {0003-4916},
doi = {https://doi.org/10.1016/0003-4916(70)90270-8},
url = {https://www.sciencedirect.com/science/article/pii/0003491670902708},
author = {Pierre Pfeuty}
}

@article{PhysRevB.35.7062,
  title = {Role of boundary conditions in the finite-size {Ising} model},
  author = {Cabrera, G. G. and Jullien, R.},
  journal = {Phys. Rev. B},
  volume = {35},
  issue = {13},
  pages = {7062--7072},
  numpages = {0},
  year = {1987},
  month = {May},
  publisher = {American Physical Society},
  doi = {10.1103/PhysRevB.35.7062},
  url = {https://link.aps.org/doi/10.1103/PhysRevB.35.7062}
}

@dataset{brodoloni_2026_18705683,
  author       = {Brodoloni, Luca and
                  Astrakharchik, Grigory and
                  Giorgini, Stefano and
                  Pilati, Sebastiano},
  title        = {Data used in ``{Energy} gap of quantum spin glasses:
                   {a} projection quantum {Monte Carlo} study''
                  },
  month        = feb,
  year         = 2026,
  publisher    = {Zenodo},
  doi          = {10.5281/zenodo.18705682}
}

@article{Venturelli2019,
  author  = {Venturelli, Davide and Kondratyev, Alexei},
  title   = {Reverse quantum annealing approach to portfolio optimization problems},
  journal = {Quantum Mach. Intell.},
  year    = {2019},
  volume  = {1},
  number  = {1},
  pages   = {17--30},
  doi     = {10.1007/s42484-019-00001-w},
  url     = {https://doi.org/10.1007/s42484-019-00001-w}
}

@article{10.3389fphy.2014.00005,
  title={{Ising} formulations of many {NP} problems},
  author={Lucas, Andrew},
  journal={Front. Phys.},
  volume={2},
  pages={74887},
  year={2014},
DOI={10.3389/fphy.2014.00005},
  publisher={Frontiers}
}

@INPROCEEDINGS{6875473,
  author={Ambainis, Andris},
  booktitle={2014 IEEE 29th Conference on Computational Complexity}, 
  title={On Physical Problems that are Slightly More Difficult than {QMA}}, 
  year={2014},
  volume={},
  number={},
  pages={32-43},
  doi={10.1109/CCC.2014.12}}

@article{Gharibian2019complexityof,
  doi = {10.22331/q-2019-09-30-189},
  url = {https://doi.org/10.22331/q-2019-09-30-189},
  title = {The complexity of simulating local measurements on quantum systems},
  author = {Gharibian, Sevag and Yirka, Justin},
  journal = {{Quantum}},
  issn = {2521-327X},
  publisher = {{Verein zur F{\"{o}}rderung des Open Access Publizierens in den Quantenwissenschaften}},
  volume = {3},
  pages = {189},
  month = sep,
  year = {2019}
}

@misc{SM,
  author       = "",
  title        = "{See Supplemental Material.}",
  howpublished = "",
  month        = "",
  year         = "",
  note         = ""
}

@article{Smith2016,
  author = {Smith, J. and Lee, A. and Richerme, P. and Neyenhuis, B. and Hess, P. W. and Hauke, P. and Heyl, M. and Huse, D. A. and Monroe, C.},
  title = {Many-body localization in a quantum simulator with programmable random disorder},
  journal = {Nat, Phys.},
  year = {2016},
  volume = {12},
  number = {10},
  pages = {907--911},
  doi = {10.1038/nphys3783},
  url = {https://doi.org/10.1038/nphys3783},
  issn = {1745-2481}
}

@article{r6dq-2ztr,
  title = {Spin-glass quantum phase transition in amorphous arrays of {Rydberg} atoms},
  author = {Brodoloni, L. and Vovrosh, J. and Juli\`a-Farr\'e, S. and Dauphin, A. and Pilati, S.},
  journal = {Phys. Rev. A},
  volume = {112},
  issue = {5},
  pages = {L051303},
  numpages = {8},
  year = {2025},
  month = {Nov},
  publisher = {American Physical Society},
  doi = {10.1103/r6dq-2ztr},
  url = {https://link.aps.org/doi/10.1103/r6dq-2ztr}
}

@article{PhysRevB.88.134204,
  title = {Typical versus averaged overlap distribution in spin glasses: {Evidence} for droplet scaling theory},
  author = {Monthus, C\'ecile and Garel, Thomas},
  journal = {Phys. Rev. B},
  volume = {88},
  issue = {13},
  pages = {134204},
  numpages = {13},
  year = {2013},
  month = {Oct},
  publisher = {American Physical Society},
  doi = {10.1103/PhysRevB.88.134204},
  url = {https://link.aps.org/doi/10.1103/PhysRevB.88.134204}
}

@Article{netket3:2022,
    title={NetKet 3: {Machine} Learning Toolbox for Many-Body Quantum Systems},
    author={Filippo Vicentini and Damian Hofmann and Attila Szabó and Dian Wu and Christopher Roth and Clemens Giuliani and Gabriel Pescia and Jannes Nys and Vladimir Vargas-Calderón and Nikita Astrakhantsev and Giuseppe Carleo},
    journal={SciPost Phys. Codebases},
    pages={7},
    year={2022},
    publisher={SciPost},
    doi={10.21468/SciPostPhysCodeb.7},
    url={https://scipost.org/10.21468/SciPostPhysCodeb.7}
}

@article{nishino2017cupy,
  title={Cupy: {A} numpy-compatible library for nvidia gpu calculations},
  author={Nishino, ROYUD and Loomis, Shohei Hido Crissman},
  journal={31st confernce on neural information processing systems},
  volume={151},
  number={7},
url = {http://learningsys.org/nips17/assets/papers/paper_16.pdf},
  year={2017}
}

@Article{10.21468/SciPostPhys.7.2.020,
	title={{QuSpin: a Python package for dynamics and exact diagonalisation of quantum many body systems. {Part II}: bosons, fermions and higher spins}},
	author={Phillip Weinberg and Marin Bukov},
	journal={SciPost Phys.},
	volume={7},
	pages={020},
	year={2019},
	publisher={SciPost},
	doi={10.21468/SciPostPhys.7.2.020},
	url={https://scipost.org/10.21468/SciPostPhys.7.2.020},
}

@article{PhysRevB.93.134202,
  title = {Effects of low-lying excitations on ground-state energy and energy gap of the {Sherrington-Kirkpatrick} model in a transverse field},
  author = {Koh, Yang Wei},
  journal = {Phys. Rev. B},
  volume = {93},
  issue = {13},
  pages = {134202},
  numpages = {14},
  year = {2016},
  month = {Apr},
  publisher = {American Physical Society},
  doi = {10.1103/PhysRevB.93.134202},
  url = {https://link.aps.org/doi/10.1103/PhysRevB.93.134202}
}

@article{PhysRevE.78.061121,
  title = {Reaching the ground state of a quantum spin glass using a zero-temperature quantum {Monte Carlo} method},
  author = {Das, Arnab and Chakrabarti, Bikas K.},
  journal = {Phys. Rev. E},
  volume = {78},
  issue = {6},
  pages = {061121},
  numpages = {6},
  year = {2008},
  month = {Dec},
  publisher = {American Physical Society},
  doi = {10.1103/PhysRevE.78.061121},
  url = {https://link.aps.org/doi/10.1103/PhysRevE.78.061121}
}

@article{PhysRevLett.70.3147,
  title = {Zero-temperature critical behavior of the infinite-range quantum {Ising} spin glass},
  author = {Miller, Jonathan and Huse, David A.},
  journal = {Phys. Rev. Lett.},
  volume = {70},
  issue = {20},
  pages = {3147--3150},
  numpages = {0},
  year = {1993},
  month = {May},
  publisher = {American Physical Society},
  doi = {10.1103/PhysRevLett.70.3147},
  url = {https://link.aps.org/doi/10.1103/PhysRevLett.70.3147}
}

@article{Bittner2006,
	doi = {10.1209/epl/i2006-10007-y},
	url = {https://doi.org/10.1209/epl/i2006-10007-y},
	year = {2006},
	month = {mar},
	publisher = {},
	volume = {74},
	number = {2},
	pages = {195},
	author = {E. Bittner and W. Janke},
	title = {Free-energy barriers in the Sherrington-Kirkpatrick model},
	journal = {Europhysics Letters}
}

@article{doi:10.1073/pnas.1910936117,
author = {Massimo Bernaschi  and Alain Billoire  and Andrea Maiorano  and Giorgio Parisi  and Federico Ricci-Tersenghi },
title = {Strong ergodicity breaking in aging of mean-field spin glasses},
journal = {Proc. Natl. Acad. Sci. U.S.A.},
volume = {117},
number = {30},
pages = {17522-17527},
year = {2020},
doi = {10.1073/pnas.1910936117},
URL = {https://www.pnas.org/doi/abs/10.1073/pnas.1910936117},
eprint = {https://www.pnas.org/doi/pdf/10.1073/pnas.1910936117}
}

@article{PhysRevE.105.034138,
  title = {Free-energy barriers in the {Sherrington-Kirkpatrick} model},
  author = {Aspelmeier, T. and Moore, M. A.},
  journal = {Phys. Rev. E},
  volume = {105},
  issue = {3},
  pages = {034138},
  numpages = {14},
  year = {2022},
  month = {Mar},
  publisher = {American Physical Society},
  doi = {10.1103/PhysRevE.105.034138},
  url = {https://link.aps.org/doi/10.1103/PhysRevE.105.034138}
}

@article{knysh2015computational,
  title={Computational bottlenecks of quantum annealing},
  author={Knysh, Sergey},
  journal={arXiv:1506.08608},
  doi={
https://doi.org/10.48550/arXiv.1506.08608},
  year={2015}
}

@article{PhysRevE.110.065305,
  title = {Zero-temperature {Monte Carlo} simulations of two-dimensional quantum spin glasses guided by neural network states},
  author = {Brodoloni, L. and Pilati, S.},
  journal = {Phys. Rev. E},
  volume = {110},
  issue = {6},
  pages = {065305},
  numpages = {9},
  year = {2024},
  month = {Dec},
  publisher = {American Physical Society},
  doi = {10.1103/PhysRevE.110.065305},
  url = {https://link.aps.org/doi/10.1103/PhysRevE.110.065305}
}

@article{PhysRevResearch.6.033322,
  title = {Finite-temperature {Rydberg} arrays: {Quantum} phases and entanglement characterization},
  author = {Reini\ifmmode \acute{c}\else \'{c}\fi{}, Nora and Jaschke, Daniel and Wanisch, Darvin and Silvi, Pietro and Montangero, Simone},
  journal = {Phys. Rev. Res.},
  volume = {6},
  issue = {3},
  pages = {033322},
  numpages = {12},
  year = {2024},
  month = {Sep},
  publisher = {American Physical Society},
  doi = {10.1103/PhysRevResearch.6.033322},
  url = {https://link.aps.org/doi/10.1103/PhysRevResearch.6.033322}
}

@article{PhysRevLett.121.167204,
  title = {Symmetries and Many-Body Excitations with Neural-Network Quantum States},
  author = {Choo, Kenny and Carleo, Giuseppe and Regnault, Nicolas and Neupert, Titus},
  journal = {Phys. Rev. Lett.},
  volume = {121},
  issue = {16},
  pages = {167204},
  numpages = {6},
  year = {2018},
  month = {Oct},
  publisher = {American Physical Society},
  doi = {10.1103/PhysRevLett.121.167204},
  url = {https://link.aps.org/doi/10.1103/PhysRevLett.121.167204}
}

@article{PhysRevE.85.036705,
  title = {Excitation gap from optimized correlation functions in quantum {Monte Carlo} simulations},
  author = {Hen, Itay},
  journal = {Phys. Rev. E},
  volume = {85},
  issue = {3},
  pages = {036705},
  numpages = {10},
  year = {2012},
  month = {Mar},
  publisher = {American Physical Society},
  doi = {10.1103/PhysRevE.85.036705},
  url = {https://link.aps.org/doi/10.1103/PhysRevE.85.036705}
}

@article{doi:10.1126/sciadv.abl6850,
author = {Markus Schmitt  and Marek M. Rams  and Jacek Dziarmaga  and Markus Heyl  and Wojciech H. Zurek },
title = {Quantum phase transition dynamics in the two-dimensional transverse-field {Ising} model},
journal = {Sci. Adv.},
volume = {8},
number = {37},
pages = {eabl6850},
year = {2022},
doi = {10.1126/sciadv.abl6850},
URL = {https://www.science.org/doi/abs/10.1126/sciadv.abl6850}
}

@article{PhysRevLett.101.170503,
  title = {Size Dependence of the Minimum Excitation Gap in the Quantum Adiabatic Algorithm},
  author = {Young, A. P. and Knysh, S. and Smelyanskiy, V. N.},
  journal = {Phys. Rev. Lett.},
  volume = {101},
  issue = {17},
  pages = {170503},
  numpages = {4},
  year = {2008},
  month = {Oct},
  publisher = {American Physical Society},
  doi = {10.1103/PhysRevLett.101.170503},
  url = {https://link.aps.org/doi/10.1103/PhysRevLett.101.170503}
}

@article{PhysRevB.94.024201,
  title = {Unconventional critical activated scaling of two-dimensional quantum spin glasses},
  author = {Matoz-Fernandez, D. A. and Rom\'a, F.},
  journal = {Phys. Rev. B},
  volume = {94},
  issue = {2},
  pages = {024201},
  numpages = {5},
  year = {2016},
  month = {Jul},
  publisher = {American Physical Society},
  doi = {10.1103/PhysRevB.94.024201},
  url = {https://link.aps.org/doi/10.1103/PhysRevB.94.024201}
}

@article{PhysRevE.87.032154,
  title = {Real-space renormalization-group approach to the random transverse-field {Ising} model in finite dimensions},
  author = {Miyazaki, Ryoji and Nishimori, Hidetoshi},
  journal = {Phys. Rev. E},
  volume = {87},
  issue = {3},
  pages = {032154},
  numpages = {10},
  year = {2013},
  month = {Mar},
  publisher = {American Physical Society},
  doi = {10.1103/PhysRevE.87.032154},
  url = {https://link.aps.org/doi/10.1103/PhysRevE.87.032154}
}

@article{PhysRevE.96.022139,
  title = {Critical and {Griffiths-McCoy} singularities in quantum {Ising} spin glasses on $d$-dimensional hypercubic lattices: {A} series expansion study},
  author = {Singh, R. R. P. and Young, A. P.},
  journal = {Phys. Rev. E},
  volume = {96},
  issue = {2},
  pages = {022139},
  numpages = {5},
  year = {2017},
  month = {Aug},
  publisher = {American Physical Society},
  doi = {10.1103/PhysRevE.96.022139},
  url = {https://link.aps.org/doi/10.1103/PhysRevE.96.022139}
}

@article{10.1561/0400000066,
    author = {Gharibian, Sevag and Huang, Yichen and Landau, Zeph and Shin, Seung Woo},
    title = {Quantum {Hamiltonian} Complexity},
    journal = {Found. Trends Theor. Comput. Sci.},
    volume = {10},
    number = {3},
    pages = {159-282},
    year = {2014},
    month = {10},
    issn = {1551-305X},
    doi = {10.1561/0400000066},
    url = {https://doi.org/10.1561/0400000066}
}

@article{Bernaschi2024,
  author = {Bernaschi, Massimo and Gonz{\'a}lez-Adalid Pemart{\'i}n, Isidoro and Mart{\'i}n-Mayor, V{\'i}ctor and Parisi, Giorgio},
  title = {The quantum transition of the two-dimensional {Ising} spin glass},
  journal = {Nature},
  year = {2024},
  volume = {631},
  number = {8022},
  pages = {749--754},
  month = {jul},
  doi = {10.1038/s41586-024-07647-y},
  url = {https://doi.org/10.1038/s41586-024-07647-y},
  issn = {1476-4687}
}

@article{RevModPhys.90.015002,
  title = {Adiabatic quantum computation},
  author = {Albash, Tameem and Lidar, Daniel A.},
  journal = {Rev. Mod. Phys.},
  volume = {90},
  issue = {1},
  pages = {015002},
  numpages = {64},
  year = {2018},
  month = {Jan},
  publisher = {American Physical Society},
  doi = {10.1103/RevModPhys.90.015002},
  url = {https://link.aps.org/doi/10.1103/RevModPhys.90.015002}
}

@article{Kazutaka2010,
doi = {10.1088/1742-6596/233/1/012008},
url = {https://doi.org/10.1088/1742-6596/233/1/012008},
year = {2010},
month = {jun},
publisher = {},
volume = {233},
number = {1},
pages = {012008},
author = {Kazutaka Takahashi and Yoshiki Matsuda},
title = {Energy-gap analysis of quantum spin-glass transitions at zero temperature},
journal = {J. Phys. Conf. Ser.}
}

@article{BAPST2013127,
title = {The quantum adiabatic algorithm applied to random optimization problems: {The} quantum spin glass perspective},
journal = {Phys. Rep.},
volume = {523},
number = {3},
pages = {127-205},
year = {2013},
issn = {0370-1573},
doi = {https://doi.org/10.1016/j.physrep.2012.10.002},
url = {https://www.sciencedirect.com/science/article/pii/S037015731200347X},
author = {V. Bapst and L. Foini and F. Krzakala and G. Semerjian and F. Zamponi}
}

@article{PRXQuantum.5.020313,
  title = {Equilibrium Dynamics of Infinite-Range Quantum Spin Glasses in a Field},
  author = {Tikhanovskaya, Maria and Sachdev, Subir and Samajdar, Rhine},
  journal = {PRX Quantum},
  volume = {5},
  issue = {2},
  pages = {020313},
  numpages = {30},
  year = {2024},
  month = {Apr},
  publisher = {American Physical Society},
  doi = {10.1103/PRXQuantum.5.020313},
  url = {https://link.aps.org/doi/10.1103/PRXQuantum.5.020313}
}

@article{PhysRevLett.129.220401,
  title = {Variational Ansatz for the Ground State of the Quantum {Sherrington-Kirkpatrick} Model},
  author = {Schindler, Paul M. and Guaita, Tommaso and Shi, Tao and Demler, Eugene and Cirac, J. Ignacio},
  journal = {Phys. Rev. Lett.},
  volume = {129},
  issue = {22},
  pages = {220401},
  numpages = {6},
  year = {2022},
  month = {Nov},
  publisher = {American Physical Society},
  doi = {10.1103/PhysRevLett.129.220401},
  url = {https://link.aps.org/doi/10.1103/PhysRevLett.129.220401}
}

@article{PhysRevB.52.3654,
  title = {Unbiased estimators in quantum {Monte Carlo} methods: Application to liquid $^{4}\mathrm{He}$},
  author = {Casulleras, J. and Boronat, J.},
  journal = {Phys. Rev. B},
  volume = {52},
  issue = {5},
  pages = {3654--3661},
  numpages = {0},
  year = {1995},
  month = {Aug},
  publisher = {American Physical Society},
  doi = {10.1103/PhysRevB.52.3654},
  url = {https://link.aps.org/doi/10.1103/PhysRevB.52.3654}
}

@article{kiss2023exact,
  title = {Complete replica solution for the transverse field {Sherrington-Kirkpatrick} spin glass model with continuous-time quantum {Monte Carlo} method},
  author = {Kiss, Annam\'aria and Zar\'and, Gergely and Lovas, Izabella},
  journal = {Phys. Rev. B},
  volume = {109},
  issue = {2},
  pages = {024431},
  numpages = {20},
  year = {2024},
  month = {Jan},
  publisher = {American Physical Society},
  doi = {10.1103/PhysRevB.109.024431},
  url = {https://link.aps.org/doi/10.1103/PhysRevB.109.024431}
}

@article{PhysRevE.96.032112,
  title = {Stability of the quantum {Sherrington-Kirkpatrick} spin glass model},
  author = {Young, A. P.},
  journal = {Phys. Rev. E},
  volume = {96},
  issue = {3},
  pages = {032112},
  numpages = {6},
  year = {2017},
  month = {Sep},
  publisher = {American Physical Society},
  doi = {10.1103/PhysRevE.96.032112},
  url= {https://link.aps.org/doi/10.1103/PhysRevE.96.032112}
}

@article{DavidLancaster1997,
doi = {10.1088/0305-4470/30/4/001},
url = {https://dx.doi.org/10.1088/0305-4470/30/4/001},
year = {1997},
month = {feb},
publisher = {},
volume = {30},
number = {4},
pages = {L41},
author = {David Lancaster and  Felix Ritort},
title = {Solving the {Schrödinger} equation for the {Sherrington} - {Kirkpatrick} model in a transverse field},
journal = {J. Phys. A Math. Gen.}
}

@article{PhysRevLett.72.4141,
  title = {Zero-temperature quantum phase transition of a two-dimensional {Ising} spin glass},
  author = {Rieger, H. and Young, A. P.},
  journal = {Phys. Rev. Lett.},
  volume = {72},
  issue = {26},
  pages = {4141--4144},
  numpages = {0},
  year = {1994},
  month = {Jun},
  publisher = {American Physical Society},
  doi = {10.1103/PhysRevLett.72.4141},
  url = {https://link.aps.org/doi/10.1103/PhysRevLett.72.4141}
}

@article{PhysRevLett.72.4137,
  title = {Quantum critical behavior of a three-dimensional {Ising} spin glass in a transverse magnetic field},
  author = {Guo, Muyu and Bhatt, R. N. and Huse, David A.},
  journal = {Phys. Rev. Lett.},
  volume = {72},
  issue = {26},
  pages = {4137--4140},
  numpages = {0},
  year = {1994},
  month = {Jun},
  publisher = {American Physical Society},
  doi = {10.1103/PhysRevLett.72.4137},
  url = {https://link.aps.org/doi/10.1103/PhysRevLett.72.4137}
}

@book{becca2017quantum,
  title={Quantum {Monte Carlo} approaches for correlated systems},
  author={Becca, Federico and Sorella, Sandro},
  year={2017},
  publisher={Cambridge University Press}
}

@article{doi:10.1126/science.aag2302,
author = {Giuseppe Carleo  and Matthias Troyer },
title = {Solving the quantum many-body problem with artificial neural networks},
journal = {Science},
volume = {355},
number = {6325},
pages = {602-606},
year = {2017},
doi = {10.1126/science.aag2302}
}

@article{CARLEO2019100311,
title = {NetKet: A machine learning toolkit for many-body quantum systems},
journal = {SoftwareX},
volume = {10},
pages = {100311},
year = {2019},
issn = {2352-7110},
doi = {https://doi.org/10.1016/j.softx.2019.100311},
url = {https://www.sciencedirect.com/science/article/pii/S2352711019300974},
author = {Giuseppe Carleo and Kenny Choo and Damian Hofmann and James E.T. Smith and Tom Westerhout and Fabien Alet and Emily J. Davis and Stavros Efthymiou and Ivan Glasser and Sheng-Hsuan Lin and Marta Mauri and Guglielmo Mazzola and Christian B. Mendl and Evert {van Nieuwenburg} and Ossian O’Reilly and Hugo Théveniaut and Giacomo Torlai and Filippo Vicentini and Alexander Wietek}
}

@article{PhysRevE.100.043301,
  title = {{Self-learning projective quantum Monte Carlo simulations guided by restricted Boltzmann machines}},
  author = {Pilati, S. and Inack, E. M. and Pieri, P.},
  journal = {Phys. Rev. E},
  volume = {100},
  issue = {4},
  pages = {043301},
  numpages = {12},
  year = {2019},
  month = {Oct},
  publisher = {American Physical Society},
  doi = {10.1103/PhysRevE.100.043301},
  url = {https://link.aps.org/doi/10.1103/PhysRevE.100.043301}
}

@article{PhysRevE.101.063308,
  title = {Simulating disordered quantum {Ising} chains via dense and sparse restricted {Boltzmann} machines},
  author = {Pilati, S. and Pieri, P.},
  journal = {Phys. Rev. E},
  volume = {101},
  issue = {6},
  pages = {063308},
  numpages = {10},
  year = {2020},
  month = {Jun},
  publisher = {American Physical Society},
  doi = {10.1103/PhysRevE.101.063308},
  url = {https://link.aps.org/doi/10.1103/PhysRevE.101.063308}
}

\clearpage

\onecolumngrid
\section{End Matter}
\twocolumngrid

\subsection{Comparing the Gaussian and binary 2D Edwards-Anderson Hamiltonians}
The energy gap is sensitive to all details of the corresponding Hamiltonian. On the other hand, properties such as the dynamical critical exponent $z$, defined from the putative power-law scaling of the disorder averaged odd gap $\left[\Delta_o\right] \propto L^{-z}$ at the critical point, is expected to depend only on the corresponding universality class and not on, \emph{e.g.}, short-range details of the Hamiltonian. This motivates us to compare the scaling obtained here for the quantum 2D-EA model with Gaussian couplings, with the corresponding results from Ref.~\cite{Bernaschi2024} for the binary case where $J_{ij}=\pm J$ with $50\%$ probability. 
\begin{figure}[h]
\centering
\includegraphics[width=1.\columnwidth]{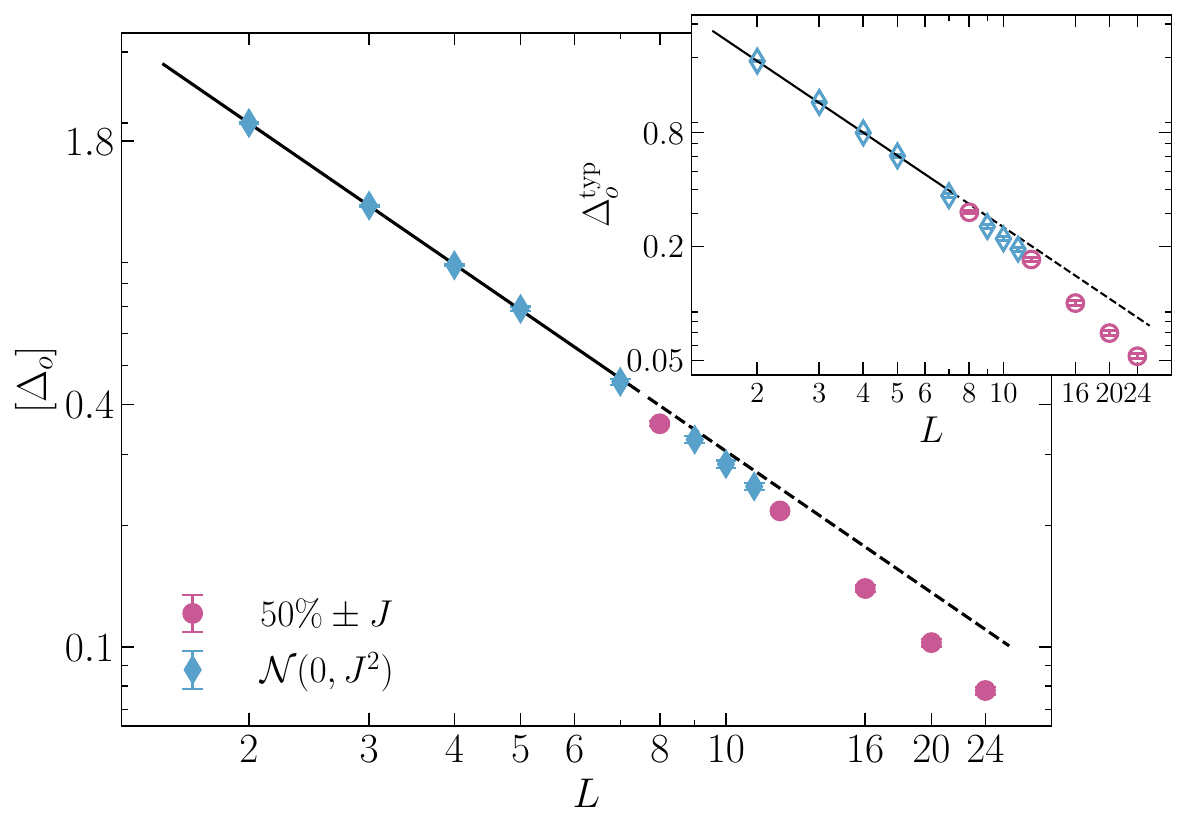}
\caption{
Disorder averaged energy odd gap $\left[\Delta_o\right]$ for the 2D-EA Hamiltonian, as a function of the lattice length $L$. The two datasets correspond to eigenvalue-solver and continuous-time PQMC calculations for Gaussian couplings $J_{ij}\propto \mathcal{N}(0,J^2)$, and to the discrete-time PIMC simulations for binary couplings $J_{ij}=\pm J$ from Ref.~\cite{Bernaschi2024}.
The deviations from the power-law fit $\left[\Delta_o\right]\propto L^{-z}$, performed on sizes up to $L=7$, 
are consistent with the super-algebraic scaling discussed in the main text.
The inset shows the scaling of the typical average $\Delta_o^{\mathrm{typ}}$.
}
\label{figEM6}
\end{figure}
As shown in Fig.~\ref{figEM6}, where an appropriate global multiplicative factor is applied to the binary-coupling results, the scalings of the two datasets are evidently consistent. At large $L$ both datasets display sizable deviations from the power-law scaling. On the one hand, the binary-coupling results, obtained through finite-temperature PIMC simulations with discrete imaginary time, reach larger system sizes. On the other hand, the continuous-time PQMC simulations are not affected by the Trotter discretization, denoting its  irrelevance for the scaling of $[\Delta_o]$. Notice that, for the binary couplings, few disorder instances were excluded -- mostly for the largest sizes -- due to an excessively small gap to be addressed via the finite temperature algorithm. The cross-validation suggests that these few gaps do not  significantly  contribute to the scaling of the disorder average.
Sizable deviations from the power-law scaling are also found considering the typical average 
$\Delta^{\mathrm{typ}}_o = \exp{\left[\ln(\Delta_o) \right]}$~\cite{PhysRevB.88.134204} (see inset of Fig.~\ref{figEM6}), where, again, square parentheses denote the average  over disorder realizations.

\begin{figure}[h]
\centering
\includegraphics[width=1.\columnwidth]{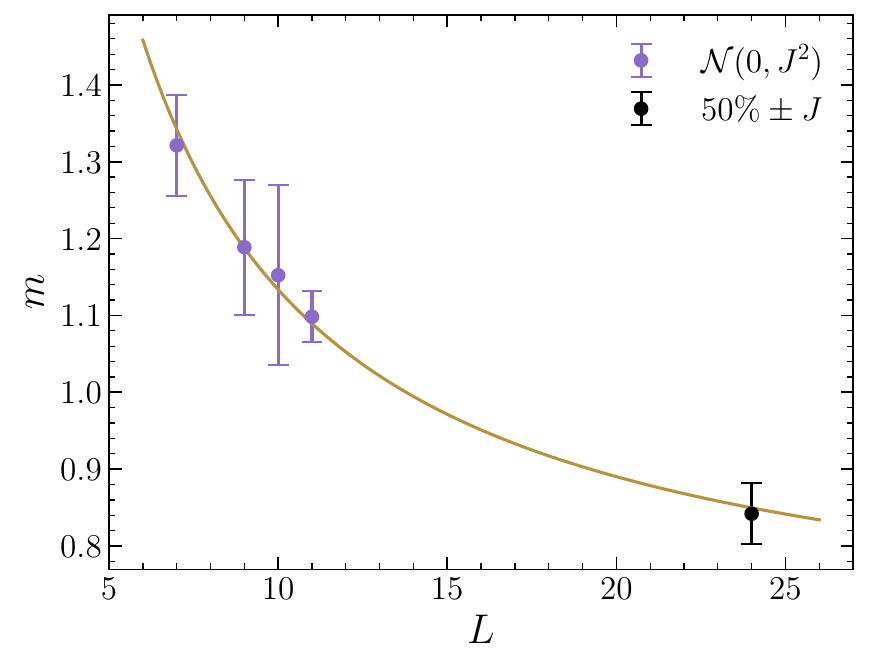}
\caption{
Power $m$ of the decay of the complementary cumulative distribution function $1-F(\eta)=B/\eta^{m}$, with $B$ and $m$ fitting parameters, as a function of the lattice length $L$ of the 2D-EA model at the spin-glass quantum phase transition. The curve is an empirical power-law fit on the Gaussian ($\mathcal{N}(0,J^2)$) results. The extrapolation agrees with the binary couplings ($50\%\pm J$) result at $L=24$.
}
\label{figEM7}
\end{figure}

To deepen the comparison between Gaussian and binary couplings, we inspect the width of the corresponding distributions of the inverse gaps $\eta=1/\Delta_o$. Following Ref.~\cite{Bernaschi2024}, a power-law fit is performed on the complementary cumulative distribution $1-F(\eta)=B/\eta^{m}$, excluding both the small $\eta$ regime and the extreme-tail regime at large $\eta$.
Specifically, the fitting range is chosen by minimizing the reduced chi-squared.
The fitted exponents $m$ for different sizes $L$ are shown in Fig.~\ref{figEM7}. The uncertainty due to the choice of fitting window is estimated by shifting its position by $10\%$ in both directions, leading to a rather conservative estimate of the error bar. By extrapolating the Gaussian result to larger lattice sizes via an empirical power-law fitting function, the binary coupling result at $L=24$ reported in Ref.~\cite{Bernaschi2024}, namely $m\simeq 0.8$, is recovered. It is worth stressing that decay rates $m < 1$ correspond to a distribution with infinite average. This analysis, which focuses on the bulk of the cumulative distribution function of $\eta$, confirms the conclusions drawn from the Hill estimator with sample fraction $\kappa=0.8$.

\subsection{Extracting the gap from PQMC imaginary-time correlations}
The PQMC simulations provide the imaginary-time correlation function
\begin{equation}
C(\tau)=\langle \hat{O}(0) \hat{O}(-\tau)\rangle\,,
\label{corr}
\end{equation}
where the brackets $\langle\;\rangle$ denote the Monte Carlo average defined in Eq.~\eqref{MCcorrelator}.
When the observable $\hat{O}$ is odd under parity symmetry, in the long imaginary-time limit $\tau\rightarrow \infty$ the correlation decays exponentially as
\begin{equation}
    C(\tau) \propto e^{-\Delta_o\tau},
\end{equation}
allowing us to extract the smallest odd gap $\Delta_o=E_{o,0}-E_0$ from an exponential fit at large $\tau$. The selection of the fitting window has to be automatized, as detailed below. 
It is worth noting that, by adopting the standard forward-walking approach~\cite{PhysRevB.52.3654}, the whole profile of the pure imaginary time correlation could be determined from
\begin{equation}
\langle\psi_0|\hat{O}(0)\hat{O}(-\tau)|\psi_0\rangle = 
\langle \hat{O}(-\Delta\tau) \hat{O}(-\Delta\tau-\tau)\rangle\,,
\end{equation}
where the time interval $\Delta\tau$ is chosen large enough to eliminate the dependence on $\psi_g(\mathbf{x})$, which would otherwise affect the behavior of the correlation function $C(\tau)$ for small values of $\tau$.

With an even-parity operator $\hat{O}$, for which the ground-state expectation value is finite, the imaginary-time correlation function converges to a finite constant at large $\tau$.
To isolate the exponential decay associated with the lowest even-parity excitation, we define the offset correlation
\begin{equation}
    \tilde{C}(\tau) = C(\tau) - \langle \hat{O}(-\tau_f)\rangle \langle \hat{O}(0)\rangle,
\end{equation}
where the imaginary time $\tau_f$ is chosen to be sufficiently long to reach the uncorrelated plateau regime.
The expectation values $\langle \hat{O}(-\tau_f)\rangle$ and $\langle \hat{O}(0)\rangle$ are independently evaluated in the PQMC simulation. The even gap $\Delta_e = E_{e,1}-E_0$ is hence extracted from an exponential fit 
$\tilde{C}(\tau)\propto e^{-\Delta_e\tau}$ in the large $\tau$ regime

\begin{figure}[h]
\centering
\includegraphics[width=1.\columnwidth]{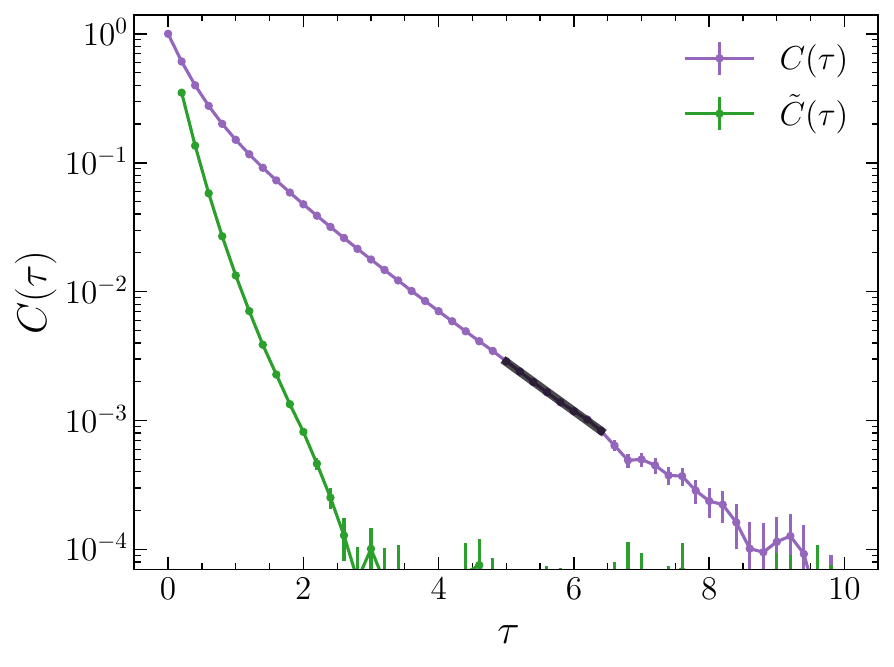}
\caption{
Odd and (offset) even imaginary-time correlation functions $C(\tau)$ and $\tilde{C}(\tau)$ for a random realization of the quantum SK model at transverse field $\Gamma=1.5$ and system size $N=100$.
The bold segment indicates the asymptotic exponential scaling from which the lowest odd $\Delta_o$ is extracted.
}
\label{figEM8}
\end{figure}

\emph{Choice of operators.}
For the odd gap we consider single-spin operators
$\hat{O}_i = \sigma_i^z.$ 
In an initial short stint of the simulation, all $N$ sites are considered, and the site index $i$ providing the largest signal at a suitably chosen imaginary time is selected on-the-fly for the remainder of the simulation.
For even parity, we consider two-site operators of the form
$\hat{O} = \sigma_i^z\sigma_j^z$. 
Again, the optimal pair $i,j$ is selected on-the-fly from a batch of $10$ randomly chosen pairs after an initial short simulation stint. 
Illustrative examples of the odd and even correlation functions for one random instance of the quantum SK model are shown in Fig.~\ref{figEM8}.

\emph{Selection of the fitting window.}
The imaginary-time correlations (${C}(\tau)$ or $\tilde{C}(\tau)$ for odd or even operators, respectively) decay exponentially only after an initial transient at small $\tau$ in which higher excitations still contribute. On the other hand, at very long imaginary times the signal-to-noise ratio deteriorates, eventually making the  signal comparable with its statistical uncertainty $\sigma(\tau)$. To isolate the region dominated by a single exponential decay, we restrict the fitting window to $\tau\in [\tau_{\mathrm{min}}, \tau_{\mathrm{max}}]$. $\tau_{\mathrm{max}}$ is tuned so to exclude data where the signal-to-noise falls below a predefined threshold, typically $C(\tau)/\sigma(\tau)\approx 8$ ($\tilde{C}(\tau)$ is used for even gaps). 
To select the left extreme $\tau_{\mathrm{min}}$, we increase it until  the fitted gap converges (from above) to a plateau value. The plateau region signals that the single-exponential assumption is valid. In practice, the procedure stops when the fitted gap remains constant within statistical uncertainty for consecutive windows or  if it starts increasing.

\clearpage
\pagestyle{plain}
\onecolumngrid
\setcounter{page}{1}
\renewcommand{\thefigure}{S\arabic{figure}}
\setcounter{figure}{0}
\begin{center}
  {\large\textbf{Supplemental material for ``Energy gap of quantum spin glasses: a projection quantum Monte Carlo study''}}

  \vspace{2em}
  { L. Brodoloni$^{1,2}$ \orcidlink{0009-0002-0887-4020}, G. E. Astrakharchik$^{3}$ \orcidlink{0000-0003-0394-8094}, S. Giorgini$^{4}$ \orcidlink{0000-0001-9146-7025}, S. Pilati$^{1, 2}$ \orcidlink{0000-0002-4845-6299}\\
  $^{1}$\small\textit{CQM group, School of Science and Technology, Physics Division, Università di Camerino, 62032 Camerino, Italy}\\
  $^{2}$\small\textit{INFN, Sezione di Perugia, I-06123 Perugia, Italy}\\
  $^{3}$\small\textit{Departament de Física, Universitat Politècnica de Catalunya, Campus Nord B4-B5, E-08034 Barcelona, Spain}\\
  $^{4}$\small\textit{Pitaevskii BEC Center, CNR-INO and Dipartimento di Fisica, Universit\`a di Trento, I-38123 Trento, Italy}\\
}
\end{center}

\vspace{0.5em}
\begin{center}
\begin{minipage}{0.8\textwidth}
\hspace{10pt} This supplemental material reports benchmark analyses on the projection quantum Monte Carlo (PQMC) algorithm for the odd and even energy gaps $\Delta_o$ and $\Delta_e$. Furthermore, the scaling with system size of the disorder-averaged inverse gap $\eta=1/\Delta$ of the quantum Sherrington-Kirkpatrick model is analyzed.
\end{minipage}
\end{center}


\maketitle

\subsection{Benchmarking the PQMC gap estimator}

Hereafter, the gap estimator discussed in the main text is tested, making a comparison against the exact results provided by alternative approaches.
A first relevant benchmark is represented by the quantum Ising chain. As extensively discussed in the literature~\cite{PFEUTY197079,PhysRevB.35.7062,PhysRevB.53.8486,10.21468/SciPostPhysLectNotes.82}, its spectrum -- including odd and even energy gaps~\cite{tang2025stretchedexponentialscalingparityrestricted} -- can be determined by applying the Jordan-Wigner transformation, which leads to a quadratic free fermion Hamiltonian. The latter can be efficiently diagonalized even for large chain lengths. Specifically, we consider a ferromagnetic chain, such that all but one nearest-neighbor couplings are equal to the energy scale $J=1$: $J_{i,i+1}=J$ for $i=1,N-1$. The remaining coupling is $J_{N,N+1}=0$, implying the use of free-end boundary conditions.
%
\begin{figure}[h!]
\centering
\includegraphics[width=0.65\columnwidth]{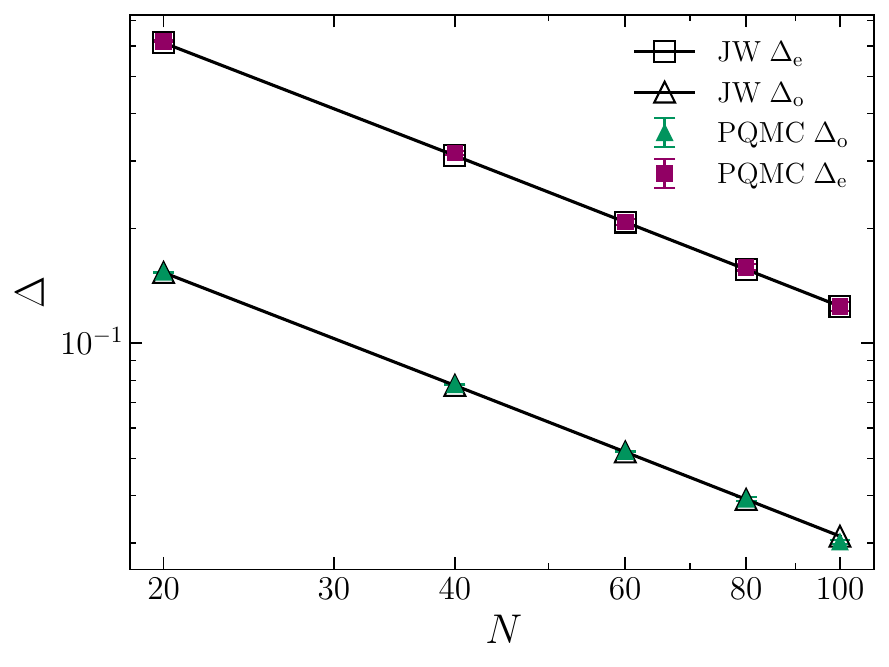}
\caption{
Energy gaps $\Delta$ of the ferromagnetic quantum Ising chain as a function of the chain length $N$. The transverse field is tuned at the paramagnet-ferromagnet quantum phase transition $\Gamma=1$.
 The odd and even gaps $\Delta_o$ and $\Delta_e$ obtained via the PQMC algorithm are represented by the solid triangles and squares, respectively. The corresponding results obtained through the Jordan-Wigner (JW) transformation are represented by hollow symbols connected by a continuous line.
}
\label{figSM1}
\end{figure}

\begin{figure}[h!]
\centering
\includegraphics[width=0.65\columnwidth]{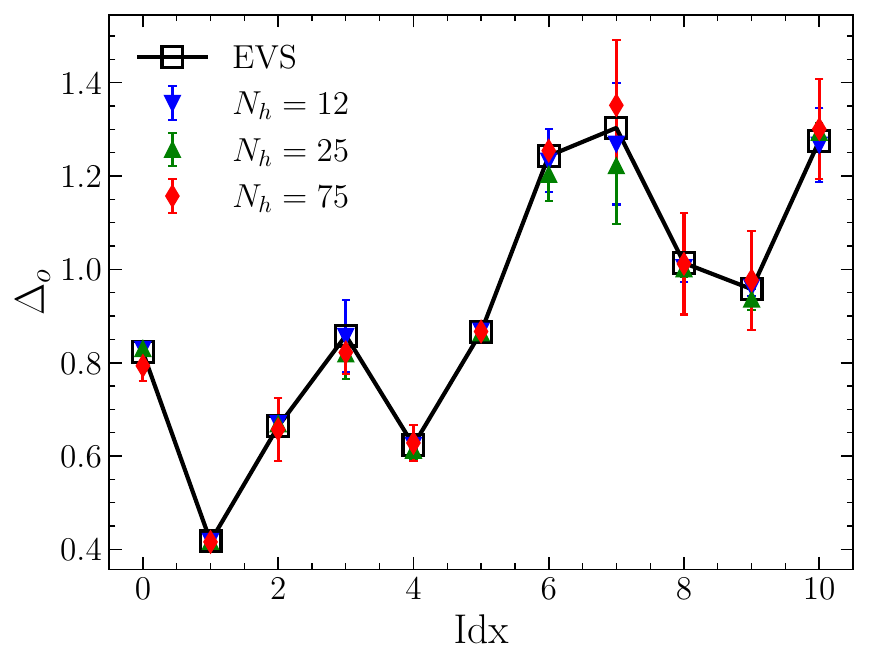}
\caption{
Odd energy gap $\Delta_o$ for 11 random realizations of the 2D Edwards-Anderson Hamiltonian. The number of spins is $N=25$, the couplings are sampled from a Gaussian distribution, and the transverse field is tuned at the spin-glass quantum critical point. The solid symbols represent the PQMC results with different guiding wave functions $\psi_g(\mathbf{x})$, featuring the number of hidden neurons $N_h$ reported in the legend. The empty squares represent the eigenvalue solver (EVS) results.
}
\label{figSM2}
\end{figure}

%
A particularly relevant testbed is represented by the quantum critical point $\Gamma=1$, which separates the paramagnetic and ferromagnetic phases.
The comparison between PQMC and Jordan-Wigner results for both the odd and even gaps is shown in Fig.~\ref{figSM1}. The precise agreement is noticeable. Both gaps are well described by the power-law scaling $\Delta\propto N^{-z}$, with the fitted dynamical critical exponent $z=0.99(1)$, in agreement with the expected exponent $z=1$~\cite{suzuki2012quantum}.

\begin{figure}[h!]
\centering
\includegraphics[width=0.65\columnwidth]{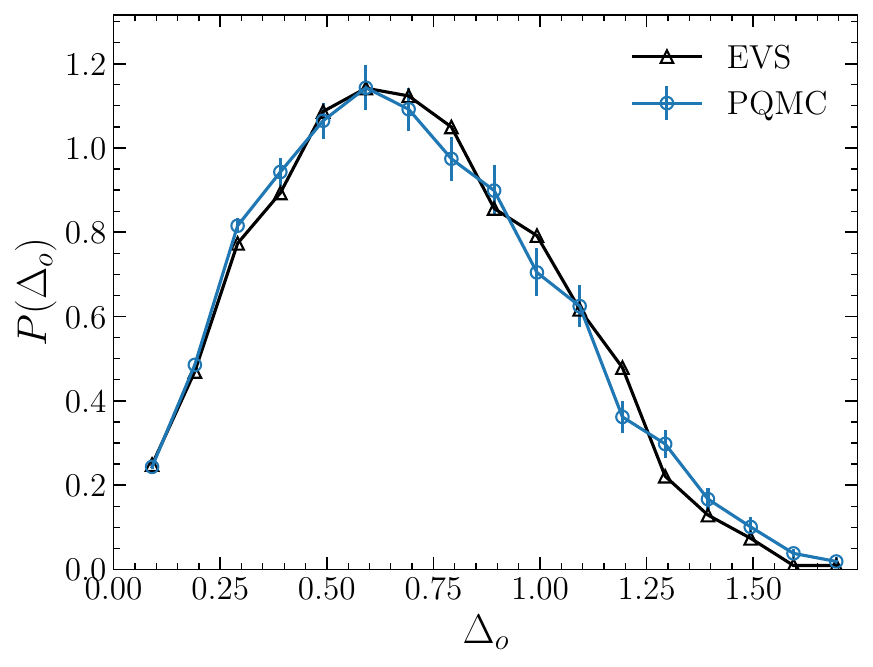}
\caption{
Histogram of the odd energy gaps $\Delta_o$ of $N_r=1000$ random realizations of the 2D Edwards-Anderson model with $N=25$ spins. The PQMC and eigenvalue-solver (EVS) results are compared.
}
\label{figSM3}
\end{figure}

\begin{figure}[h!]
\centering
\includegraphics[width=0.65\columnwidth]{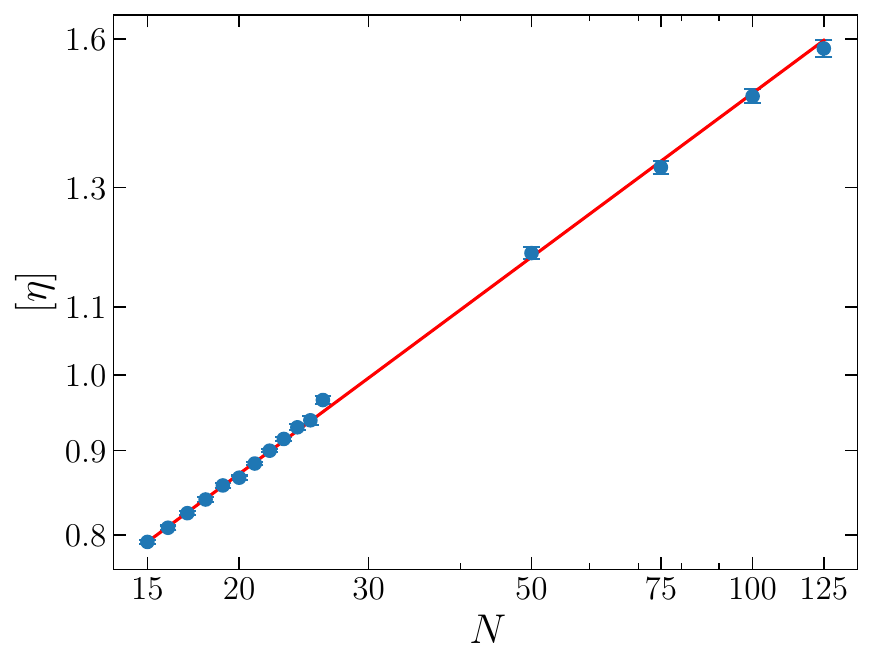}
\caption{
Disorder average $[\eta]$ of the inverse (odd) gap $\eta=1/\Delta$ as a function of the number of spins $N$ for the quantum SK model at the spin-glass quantum phase transition. The red line represents the power-law scaling $[\eta]\propto N^{1/3}$.
}
\label{figSM4}
\end{figure}

Beyond the one-dimensional case, exact theoretical approaches are not available. Hence, we benchmark the PQMC algorithm against gaps numerically obtained via the exact eigenvalue solver described in the main text. 
Specifically, we consider the two-dimensional Gaussian Edwards-Anderson model of size $N=L^2=25$ at the transverse field  $\Gamma=1.98$, corresponding to the estimate of the critical point $\Gamma=1.98(7)$~\cite{PhysRevE.110.065305}. Fig.~\ref{figSM2} shows the odd gaps for 11 realizations of the disorder. These gaps display large random fluctuations. Yet, the predictions of the PQMC algorithm match the exact results within the statistical uncertainties for all instances.
Notably, the PQMC results are independent of the number of hidden neurons $N_h$ included in the restricted Boltzmann machines. This finding numerically confirms the derivation provided in the main text, which asserts that the PQMC gap estimator is independent of the choice of the guiding wave function $\psi_g(\mathbf{x})$.

To further assess the reliability of the PQMC gap estimator at scale, we analyze the odd gap distribution across a large ensemble of disorder realizations. The testbed model is, again, the 2D critical Gaussian EA Hamiltonian including $N=25$ spins. The comparison of the gap histogram against the corresponding eigenvalue-solver results -- shown in Fig.~\ref{figSM3} -- denotes, again, the fidelity of the PQMC estimator.
The error bars of the PQMC histogram are determined by averaging over a large ensemble of randomized gap datasets with added Gaussian fluctuations scaled with the statistical uncertainty of the corresponding gap values.
%

\subsection{Scaling of the inverse gap}
The scaling of the disorder-averaged fundamental gap $[\Delta]$ (we identified $\Delta \equiv \Delta_o$) for the quantum SK model is analyzed in the main text. Hereafter, we consider instead the scaling of the disorder average $[\eta]$ of the inverse gap $\eta=1/\Delta$. As shown in Fig.~\ref{figSM4}, the scaling with system size $N$ is well approximated by a power-law function $[\eta]\propto N^{0.33}$. 
Notice that this exponent turns out to agree with the scaling of the inverse of the disorder-averaged gap.


\end{document}